\documentclass[pra,twocolumn,showpacs]{revtex4}
\usepackage{amsmath,amssymb}
\usepackage{bm}
\usepackage{psfrag}
\usepackage{graphicx}
\newcommand{\op}[1]{\hat{#1}}
\newcommand{\infint}{\int_{-\infty}^\infty}
\newcommand{\ii}{i}
\newcommand{\dd}{d}
\newcommand{\ee}{e}
\newcommand{\NN}{\nonumber\\}
\newcommand{\sub}[1]{_{\mbox{\scriptsize#1}}}

\def\ket#1{|#1\rangle}
\def\bra#1{\langle #1|}
\def\bracket#1{\langle #1 \rangle}
\def\bracketi#1#2{\langle #1|#2 \rangle}

\begin{document}
\title{Is the Heisenberg uncertainty relation 
really violated?}
% Defense of the Heisenberg uncertainty relation
\author{Masao Kitano}
\email{kitano@kuee.kyoyo-u.ac.jp}
\affiliation{Department of Electronic Science and Engineering,
Kyoto University, Kyoto 615-8510, Japan}
\affiliation{CREST, Japan Science and Technology Agency, Tokyo 102-0075, Japan}
\date{\today}

\begin{abstract}
It has been pointed out that
for some types of measurement
the Heisenberg uncertainty relation seems to be violated.
In order to save the situation a new uncertainty relation was proposed by Ozawa.
Here we introduce revised definitions of error and disturbance
taking into account the gain associated with generalized measurement interactions.
With these new definitions, the validity of the Heisenberg inequality 
is recovered for continuous linear measurement interactions.
We also examine the changes in distribution functions caused by the
general measurement interaction and clarify
the physical meanings of infinitely large errors and disturbances.
\end{abstract}



\maketitle

\section{Introduction}
The uncertainty relation for quantum states
is that 
the fluctuations for the canonically conjugate
observables $\op q$ and $\op p$
must satisfy
the relation
$\sigma(\op q)\sigma(\op p)\geq\hbar/2$.
This can easily be proven using
the canonical commutation relation
$[\op q, \op p]=\ii\hbar\op1$ \cite{kennard,peres}.
It is closely connected with the complementarity
or the wave-particle duality of quantum states.
Practically it places limitations in the
preparation of wave packets.

On the other hand, the uncertainty relation
that was discussed by Heisenberg earlier
in terms of the hypothetical gamma-ray microscope is
for the quantum measurement \cite{heisenberg,heisenberg-chicago}.
Namely, if one measures an observable $\op q$
with a given accuracy (or with error) $\epsilon$,
then the conjugate observable $\op p$ necessarily
suffers the disturbance $\eta$, which satisfies
the inequality 
\begin{align}
\epsilon\eta\geq\frac{\hbar}{2}
.
\label{eq2}
\end{align}

The Heisenberg inequality can be derived by utilizing
a quantum mechanical model of measurement
processes \cite{neumann}.
As the first step of measurement the unitary
interaction between the object and a probe of
measuring apparatus is considered.
The probe is a part of measuring apparatus and 
works as a front end that interacts
with the object quantum-mechanically.
After the interaction, the probe variable $\op Q$,
which contains some information on the object
variable $\op q$,
is read out by the other part of the measuring
apparatus and is fixed as a classical value.
This indirect measurement model is helpful
to discuss the error and disturbance
associated with measurement.

It has been pointed out that
for some types of interactions
the Heisenberg inequality (\ref{eq2}) seems to be violated
\cite{ozawa-prl,caves}.
For example, in the contractive measurement \cite{yuen},
$\epsilon=0$ and $\eta<\infty$ are satisfied simultaneously and
the inequality is violated; $\epsilon\eta=0$.
The other extreme case of $\eta=0$ and $\epsilon<\infty$ is
also possible.

In order to reconcile with the above cases,
a new uncertainty relation was proposed
by Ozawa \cite{ozawa-pl,ozawa-pra,ozawa-ap}:
\begin{align}
\epsilon\eta+\epsilon\sigma(\op p)+\sigma(\op q)\eta\geq\frac{\hbar}{2}    
.
\label{eq4}
\end{align}

The Ozawa inequality is attracting considerable interests
because it admits the cases that violate the limit
posed by the Heisenberg inequality \cite{hall,busch,koshino,kurotani}.
It might be possible to devise a measurement scheme
which is free from the Heisenberg uncertainty principle.
Unfortunately, however, no systematic experiments that
demonstrate the violation of the Heisenberg limit
or that make some use of the new bound have been performed so far.

From experimentalists' view point, the definition of
the noise and disturbance operators,
whose expectation values give the error and 
disturbance, is the first obstacle.
Each of them contains two observables,
one for before and the other for after the interaction.
Moreover because they are non-commutable,
the separate measurements are of no use and a joint measurement
seems required essentially \cite{werner,arthurs-kelly}.
However, this difficulty is only an artifact associated with the use of
the Heisenberg picture in the theory.
It will be shown that the use of Schr\"odinger picture 
gives more simple perspective.
With the probability distributions 
of corresponding observable, separately measured before
and after the interaction, we can examine the error and
disturbance.

Yet there seems another problem in the definition of
the noise and disturbance operators.
In the derivation of the Ozawa inequality (\ref{eq4}), a broad class of
measurement interaction is assumed.
On the other hand, the definition of the noise and disturbance
operators are just borrowed from the case of ideal measurement.
In the case of general interactions, we have to consider the
amplification (or deamplification) of variables due to the
unitary transformation \cite{arthurs-goodman,gavish}.
The gain, which is unity for the ideal measurement, must be
taken into account.
We will show that redefining the error and disturbance operators
properly, the violation of the Heisenberg inequality is canceled for
a broad class of interaction.

A related problem 
in the discussion of violation of the Heisenberg inequality
is that the following is admitted unconditionally:
the finiteness of disturbance $\eta$ is
implied from the finite standard deviation $\sigma(\op p')$ of
the object momentum after the interaction.
Even for the case of finite standard deviation, if the 
distribution of $\op p'$ is completely
uncorrelated with the original distribution of
$\op p$, then the disturbance should be
considered infinitely large so as to destroy
the information completely.
Similarly the finiteness of the variance $\sigma(\op Q')$ of
the probe position after the interaction
does not imply the finite error $\epsilon$.
The error should be considered infinite if the information on $\op q$ is not
transferred to $\op Q'$ at all.
These claims will be confirmed
by examining the change of probability distributions
with the Schr\"odinger picture (Sec.~VII).

\section{The uncertainty relation for the standard model}
The measurement process can be described with
the object system to be measured
and the probe.
The probe is the front end of the measuring device
and is assumed to be treated quantum-mechanically.
The measured observable (position) is denoted by $\op q$ and its conjugate
observable (momentum) by $\op p$.
For the probe, the conjugate observables $\op Q$ and $\op P$ are
introduced.
We assume, $[\op q,\op p]=[\op Q,\op P]=\ii\hbar$ and
the eigenkets for $\op q$ and $\op Q$ are defined with
$\op q\ket q=q\ket q$ and
$\op Q\ket Q=Q\ket Q$, respectively.

The probe is prepared in a fixed known state $\ket\varPsi$ and the object is 
of course in an unknown state $\ket\psi$.
The two systems are made interacted for a given period of time.
The interaction can be represented with a unitary operator
$\op U$.
Then the probe variable $\op Q$ is measured by the next
stage of the measuring apparatus.
This part can be modeled with von Neumann type (projection) measurement.

For the moment,
we assume that the unitary operator $\op U$ satisfies the relation
\cite{braginsky},
\begin{align}
\op Q' = \op U^\dagger\op Q\op U=\op Q + \op q,\quad
\op p' = \op U^\dagger\op p\op U=\op p - \op P    
\label{eq10}
\end{align}
where $\op Q'$ and $\op p'$ are the quantities after the interaction
(in the Heisenberg picture).
This interaction corresponds to the case of ideal measurement.

Rewriting (\ref{eq10}) as
\begin{align}
\op Q'
=\op q + \op E
,\quad
\op p'
=\op p + \op D,
\label{eq20}
\end{align}
we find the definition of
two operators
\begin{align}
\op E := \op Q' - \op q\,(=\op Q)
,\quad
\op D := \op p' - \op p\,(=-\op P)
.
\label{eq25}
\end{align}
The former operator corresponds to the accuracy or the error
added to $\op q$ and the latter operator corresponds to the
disturbance against $\op p$.

The second-order moments 
of $\op E$ and $\op D$ 
for an initial state $\ket{\varPsi\sub{tot}}=\ket\psi\ket\varPsi$ give
the error $\epsilon$ and $\eta$ as
\begin{align}
\epsilon := \bracket{\op E^2}^{1/2} \geq \sigma(\op Q)
,\quad
\eta := \bracket{\op D^2}^{1/2} \geq \sigma(\op P)    
,
\label{eq30}
\end{align}
respectively.
For an operator $\op A$,
$\bracket{\op A}$ represents the expectation value and
$\sigma(\op A)=(\bracket{\op A^2}-\bracket{\op A}^2)^{1/2}$ is
the standard deviation with
respect to a given state.
Hereafter, for simplicity, we assume that the initial state $\ket\varPsi$ of
probe satisfies the
conditions $\bracket{\op Q}=\bracket{\op P}=0$.
Then the equalities hold in Eq.~(\ref{eq30}),
from which Eq.~(\ref{eq30}) we have the
Heisenberg uncertainty relation (HUR)
\begin{align}
\epsilon\eta \geq \sigma(\op Q)\sigma(\op P) \geq \frac{\hbar}{2}
\label{eq40}
\end{align}
for the indirect measurements with ideal interaction (\ref{eq10}).

\section{Interaction for measurement of continuous variables}
The interaction (\ref{eq10}) is for the ideal
measurement of a continuous variable.
Using the eigenstates for the positions of object and probe,
the action of $\op U$ can be written as
\begin{align}
\op U:\,
\ket q\ket Q \mapsto \ket q\ket{Q+q}
.
\label{eq50}
\end{align}
The probe position $Q$ is deflected by the variable 
$q$ to be measured, while $q$ itself is not affected by
the interaction.

Here we introduce a generalized form of interaction $\op U$:
for $a, b, c, d\in \mathbb{R}$, it is defined as
\begin{align}
\op U:\,
\ket q\ket Q \mapsto \sqrt{\varDelta}\ket{dq+cQ}\ket{aQ+bq}
.
\label{eq60}
\end{align}
The positions ($\op q$, $\op Q$) are linearly transformed 
through the interaction \cite{braginsky}.
The factor $\sqrt\varDelta$ with
$\varDelta=ad-bc>0$ is determined from the unitary condition
of $\op U$ (Appendix).

With this unitary transformation, the variables for the object and probe are transformed 
\begin{align}
\begin{bmatrix}
\op Q' \\ \op q'        
\end{bmatrix}
=
\begin{bmatrix}
a & b \\ c & d    
\end{bmatrix}
\begin{bmatrix}
\op Q \\ \op q    
\end{bmatrix}
,\quad
\begin{bmatrix}
\op p' \\ \op P'        
\end{bmatrix}
=
\frac{1}{\varDelta}
\begin{bmatrix}
a & -b \\ -c & d    
\end{bmatrix}
\begin{bmatrix}
\op p \\ \op P    
\end{bmatrix}
,
\label{eq140}
\end{align}
where
$\op Q' = \op U^\dagger\op Q\op U$,
$\op q' = \op U^\dagger\op q\op U$,
$\op p' = \op U^\dagger\op p\op U$,
and
$\op P' = \op U^\dagger\op P\op U$.
Corresponding to Eq.~(\ref{eq10}),
the portions related to the uncertainty relation are
\begin{align}
\op Q' = a\op Q + b\op q,
\quad
\op p' = a'\op p - b'\op P,    
\label{eq150}
\end{align}
where $a'=a/\varDelta$, $b'=b/\varDelta$.
The parameters $a$, $b$, and $\varDelta$ are relevant to 
the uncertainty relation.
We can assume $a\geq0$ and $b\geq0$ as discussed in Appendix.
The parameter $b$ corresponds to the gain from $\op q$ to $\op Q'$
and $a'$ that from $\op p$ to $\op p'$.

\section{Standard forms of interaction}
We can classify the unitary transformation
(\ref{eq140}) into three types and associate a standard
form to each class.

Before and after the interaction, we apply the following scale transformations 
\begin{align}
&\op Q \rightarrow\varLambda \op Q,\quad
\op P \rightarrow\varLambda^{-1}\op P,\quad
\op q \rightarrow\lambda\op q,\quad
\op p \rightarrow\lambda^{-1}\op p,
\NN
&\op q' \rightarrow\mu^{-1} \op q',\quad
\op p' \rightarrow\mu\op p',
\label{eq160}
\end{align}
where $\varLambda$, $\lambda$, and $\mu$ are non-zero, real constants.
Then the coefficient matrix transforms as
\begin{align}
\begin{bmatrix}
a & b \\ c & d    
\end{bmatrix}
\rightarrow
\begin{bmatrix}
1 & 0 \\ 0 & \mu        
\end{bmatrix}
\begin{bmatrix}
a & b \\ c & d    
\end{bmatrix}
\begin{bmatrix}
\Lambda & 0 \\ 0 & \lambda        
\end{bmatrix}
=
\begin{bmatrix}
\varLambda a & \lambda b\\
\mu\varLambda c & \mu\lambda d
\end{bmatrix}
.
\label{eq165}
\end{align}
In the case of $ab\neq0$,
by setting
$\varLambda=1/a$,
$\lambda=1/b$,
$\mu=ab/\varDelta$,
we can simplify the matrices as
\begin{align}
\begin{bmatrix}
\op Q' \\ \op q'        
\end{bmatrix}
=
\begin{bmatrix}
1 & 1 \\
b'c & a'd
\end{bmatrix}
\begin{bmatrix}
\op Q \\ \op q    
\end{bmatrix}
,\quad
\begin{bmatrix}
\op p' \\ \op P'        
\end{bmatrix}
=
\begin{bmatrix}
1 & -1\\
-b'c & a'd
\end{bmatrix}
\begin{bmatrix}
\op p \\ \op P    
\end{bmatrix}
\label{eq180}
.
\end{align}

In the case $a=0$, $b\neq0$,
we can set
$\lambda=1/b$, $\varLambda=-1/c$, $\mu=1$,
to obtain
\begin{align}
\begin{bmatrix}
\op Q' \\ \op q'        
\end{bmatrix}
=
\begin{bmatrix}
0 & 1 \\
-1 & d/b
\end{bmatrix}
\begin{bmatrix}
\op Q \\ \op q    
\end{bmatrix}
,\quad
\begin{bmatrix}
\op p' \\ \op P'        
\end{bmatrix}
=
\begin{bmatrix}
0 & -1\\
1 & d/b
\end{bmatrix}
\begin{bmatrix}
\op p \\ \op P    
\end{bmatrix}
.
\label{eq190}
\end{align}

Similarly in the case $a\neq0$, $b=0$,
by setting
$\lambda=1/d$, $\varLambda=1/a$, $\mu=1$,
we have
\begin{align}
\begin{bmatrix}
\op Q' \\ \op q'        
\end{bmatrix}
=
\begin{bmatrix}
1 & 0 \\
c/a & 1
\end{bmatrix}
\begin{bmatrix}
\op Q \\ \op q    
\end{bmatrix}
,\quad
\begin{bmatrix}
\op p' \\ \op P'        
\end{bmatrix}
=
\begin{bmatrix}
1 & 0\\
-c/a & 1
\end{bmatrix}
\begin{bmatrix}
\op p \\ \op P    
\end{bmatrix}
.
\label{eq200}
\end{align}

Now we have the three standard forms for measurement
interaction;
\begin{align}
\begin{array}{cllll}
&\text{(O)}\quad
\op Q'=\op Q + \op q,
\quad
\op p'=\op p - \op P,
\\
&\text{(A)}\quad
\op Q'=\op q,
\quad
\op p'=-\op P,
\\
&\text{(B)}\quad
\op Q'=\op Q,
\quad
\op p'=\op p
\end{array}
.
\label{eq210}
\end{align}

Type (O), to which the ideal measurement ($c=0$) belongs, covers a wide class
of interactions, $ab\neq0$.
According to the definitions of error and disturbance, i.e.,
Eqs.~(\ref{eq25}) and (\ref{eq30}), we have
\begin{align}
&\op E=\op Q,\quad \op D=-\op P,
\NN
&\epsilon \geq \sigma(\op Q),\quad
\eta\geq \sigma(\op P)
\label{eq220}
\end{align}
and
the uncertainty relation
can easily been derived;
 $\epsilon\eta=\sigma(\op Q)\sigma(\op P)\geq\hbar/2$.

For Type (A), we have
\begin{align}
&\op E=0,\quad \op D=-\op P-\op p,
\NN
&\epsilon = 0,\quad
\eta\geq \sqrt{\sigma^2(\op P)+\sigma^2(\op p)}
.
\label{eq230}
\end{align}
The probe variable $\op Q'$ after the interaction
turns into the variable $\op q$ itself, therefore,
no errors come in.
On the other hand, the disturbance can be finite.
The case of swapping interaction $(d=0)$ and the contractive
interaction \cite{yuen} $(d/b=1)$ are contained in this 
class.

For Type (B), we have
\begin{align}
&\op E=\op Q - \op q,\quad \op D=0,
\NN
&\epsilon\geq \sqrt{\sigma^2(\op Q)+\sigma^2(\op q)},\quad
\eta= 0
.
\label{eq240}
\end{align}
The conjugate variable is conserved: $\op p'=\op p$, therefore,
no disturbances occur, while the
error can be finite.
This type of interaction is singular
in the sense that the probe variable $\op Q'$ does not
depend on the variable $\op q$ to be measured.
The case of no-interaction ($c=0$) is contained
in this type.
The case of ideal measurement for the object's momentum
($c/a=-1$) is also contained.

\section{Ozawa's inequality}
We have seen that the
Heisenberg uncertainty relation holds for 
Type (O) interaction but seems broken for
Types (A) and (B),
actually, $\epsilon\eta=0$ for these cases.

A new inequality which is valid for all types has been
proposed by Ozawa \cite{ozawa-pl}:
\begin{align}
\epsilon\eta+\epsilon\sigma(\op p)+\sigma(\op q)\eta\geq\frac{\hbar}{2}
\label{eq250}
\end{align}
The Ozawa uncertainty relation (OUR) 
includes the standard deviations
$\sigma(\op q)$ and $\sigma(\op p)$ of the initial
object state in addition to $\epsilon$ and $\eta$.

For Type (A), 
OUR sets a finite lower bound of disturbance
as $\eta\geq(\hbar/2)\sigma(\op q)^{-1}\geq\sigma(\op p)$ despite of
$\epsilon=0$.
For type (B),
the finite error $\epsilon\geq\sigma(\op q)$ for $\eta=0$.

\begin{figure}
\psfrag{x}[c][c]{$\tilde{\epsilon}=\epsilon/\sigma(q)$}
\psfrag{y}[c][c]{$\tilde{\eta}=\eta/\sigma(p)$}
\psfrag{1}{$1$}
\psfrag{0}{$0$}
\psfrag{OUR}{OUR}
\psfrag{HUR}{HUR}
\psfrag{a=0.01}{$a=0.01$}
\psfrag{b=0.01}{$b=0.01$}
\centering
\includegraphics[scale=0.7]{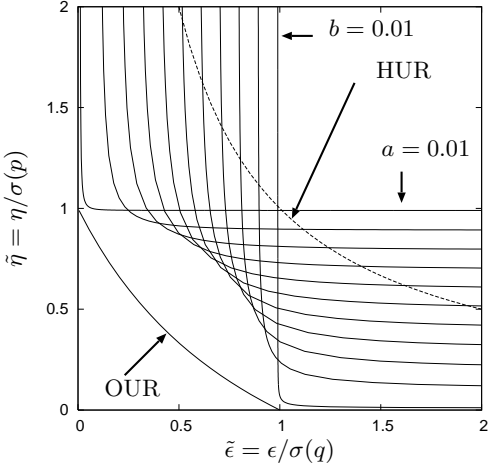}    
\caption{
The Heisenberg limit $\tilde{\epsilon}\tilde{\eta}=1$ (HUR, dashed line)
and the Ozawa limit 
$\tilde{\epsilon}\tilde{\eta}+\tilde{\epsilon}+\tilde{\eta}=1$ (OUR, real line) 
are plotted on the $(\tilde\epsilon,\tilde\eta)$-plane.
The trajectories of the
normalized uncertainties (\ref{eq300}) for general interactions are 
plotted for some parameters
($a+b=0$, $\varDelta=1$, $a=0.01, 0.1, 0.2, \cdots, 0.9, 0.99$).
}
\label{fig100}
\end{figure}

We explore the relation between HUR and OUR\@.
Assuming the minimum uncertainty for the initial state of object;
$\sigma(\op q)\sigma(\op p)=\hbar/2$,
we can introduce normalized variables
$\tilde{\epsilon}=\epsilon/\sigma(\op q)$ and $\tilde{\eta}=\eta/\sigma(\op p)$.
The normalization gives
\begin{align}
& \tilde{\epsilon}\tilde{\eta}\geq 1 &\text{(HUR)},
\\    
& \tilde{\epsilon}\tilde{\eta} + \tilde{\epsilon} + \tilde{\eta} \geq 1 &\text{(OUR)}.
\label{eq260}
\end{align}
As shown in Fig.~\ref{fig100}, 
there is an appreciable gap between the two bounds and
the bound of OUR always violates HUR.

Using Eq.~(\ref{eq25}), the error and disturbance operators
$\op E$ and $\op D$ 
for the general
interaction (\ref{eq150}) are
\begin{align}
\op E = a\op Q + (b-1)\op q
,\quad
\op D = (a'-1)\op p - b'\op P
.
\label{eq270}
\end{align}
The expectation values of
$\op E^2$ and $\op D^2$ for an initial state are
\begin{align}
\epsilon^2 &= \bracket{\op E^2} = 
a^2\sigma^2(\op Q)+(b-1)^2\sigma^2(\op q)
\NN
\eta^2 &= \bracket{\op D^2} = 
(a'-1)^2\sigma^2(\op p)+b'^2\sigma^2(\op P)
.
\label{eq280}
\end{align}
For simplicity we have assumed $\bracket{\op q}=\bracket{\op p}=0$.
(By introducing new variables:
$\op q_1=\op q - \bracket{\op q}$,
$\op Q_1'=\op Q' - a\bracket{\op q}$,
$\op p_1=\op p - \bracket{\op p}$,
$\op p_1'=\op p' - a'\bracket{\op p}$,
we can remove the mean values.)

We introduce a parameter
\begin{align}
w=\sigma(\op Q)/\sigma(\op q)=\sigma(\op p)/\sigma(\op P) >0
,
\label{eq290}
\end{align}
which determines the balance of variances of the object and probe variables.
A minimum uncertainty states for the object and probe are assumed;
$\sigma(\op q)\sigma(\op p)=\sigma(\op Q)\sigma(\op P)=\hbar/2$.
With these, Eq.~(\ref{eq280}) can be written as
\begin{align}
\tilde{\epsilon}^2=a^2 w^2 + (b-1)^2,
\quad
\tilde{\eta}^2=(a'-1)^2 + b'^2 w^{-2}    
.
\label{eq300}
\end{align}
In Fig.~\ref{fig100},
several trajectories
$\{(\tilde{\epsilon}(w),\tilde{\eta}(w))\,|\, 0< w<\infty\}$ 
are plotted for some combinations of parameters,
$a$, $b$, and $\varDelta$.
We see that HUR is violated for some cases while
OUR is respected all the time.

For the special case $a'=b=1$, or for the standard form
(O), the trajectory coincides with the Heisenberg limit because of
$\tilde{\epsilon}^2=\varDelta^2w^2$, $\tilde{\eta}^2=\varDelta^{-2}w^{-2}$.
This is the only case where the HUR is valid irrespective of the
value $w$.

We notice that at least within the framework of linear interaction (\ref{eq150}),
we can have a bound tighter than that of OUR (but weaker
than HUR),
namely,
\begin{align}
\tilde{\epsilon}^2+\tilde{\eta}^2
&=(a'-1)^2+(b-1)^2+a^2w^2+b'^2w^{-2}
\NN
&\geq(a'-1)^2+(b-1)^2+2a'b
\NN
&=(a'+b-1)^2+1\geq1    
.
\label{eq310}
\end{align}
The bound corresponds to the (quarter) circle of unit radius centered at the origin,
which is seen as an envelope in Fig.~\ref{fig100}.

The Ozawa inequality has been introduced as a remedy
against the violation of the Heisengberg inequality
in the cases of (A) and (B), which represent somewhat
singular interactions.

\begin{figure}[t]
\psfrag{(a)}{(a)}
\psfrag{(b)}{(b)}
\psfrag{(c)}{(c)}
\psfrag{a}[c][c]{$a$}
\psfrag{b}[c][c]{$b$}
\psfrag{1}[c][c]{$1$}
\psfrag{b-1}[c][c]{$b-1$}
\psfrag{1/b}[c][c]{$1/b$}
\psfrag{q}[c][c]{$\op q$}
\psfrag{Q'}[c][c]{$\op Q'$}
\psfrag{Q}[c][c]{$\op Q$}
\psfrag{E}[c][c]{$\op E$}
\psfrag{E*}[c][c]{$\op E_*$}
\centering
\includegraphics[scale=0.8]{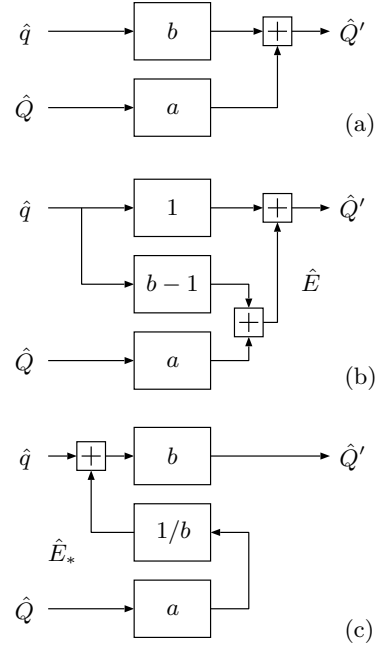}    
\caption{
Graphical relations of position operators in measurement interaction;
(a)
$\op Q' = a\op Q + b\op q$,
(b) 
$\op Q' = \op q + \op E$,
(c)
$\op Q' = b(\op q + \op E_*)$.
}
\label{fig200}
\end{figure}

\begin{figure}[t]
\psfrag{(a)}{(a)}
\psfrag{(b)}{(b)}
\psfrag{(c)}{(c)}
\psfrag{a'}[c][c]{$a'$}
\psfrag{-b'}[c][c]{$-b'$}
\psfrag{1}[c][c]{$1$}
\psfrag{a'-1}[c][c]{$a'-1$}
\psfrag{1/a'}[c][c]{$1/a'$}
\psfrag{p}[c][c]{$\op p$}
\psfrag{p'}[c][c]{$\op p'$}
\psfrag{P}[c][c]{$\op P$}
\psfrag{D}[c][c]{$\op D$}
\psfrag{D*}[c][c]{$\op D_*$}
\centering
\includegraphics[scale=0.8]{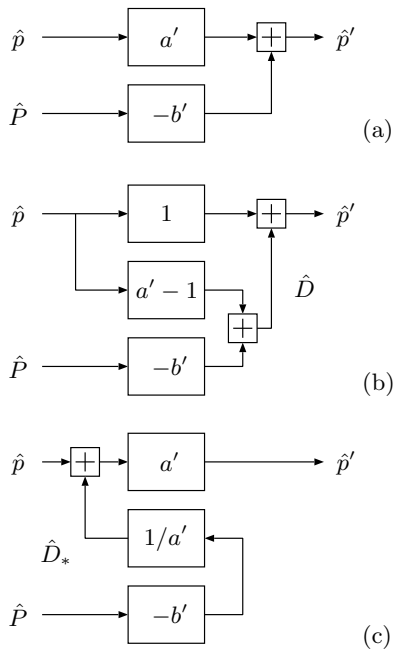}    
\caption{
Graphical relations of momentum operators in measurement interaction;
(a)
$\op p' = a'\op p - b'\op P$,
(b) 
$\op p' = \op p + \op D$,
(c)
$\op p' = a'(\op p + \op D_*)$.
}
\label{fig300}
\end{figure}

\section{Redefinition of noise and disturbance}
We propose another method of remedy for the violation of
the Heisenberg uncertainty relation.
We replace the definition of noise and disturbance operators
considering the gain of the interaction.

We use the general (unscaled) interaction
(\ref{eq150}) with the parameters $a$, $b$, and $\varDelta$.
For the moment, $ab\neq0$ is assumed.
Equation (\ref{eq150}) can be rewritten as
\begin{align}
\op Q'
=b(\op q + \op E_*)
,\quad
\op p'
=a'(\op p + \op D_*)
,
\label{eq320}
\end{align}
where 
$b$ can be considered as the gain
with which the input $\op q$ is amplified to generate the output $\op Q'$.
Similarly 
$a'=a/\varDelta$ is the gain from $\op p$ to $\op p'$.
Here, we have replaced the definition of $\op E$ and $\op D$ by
\begin{align}
&\op E_* := \frac{1}{b}\op Q' - \op q = \frac{a}{b}\op Q
,
\NN
&\op D_* := \frac{1}{a'}\op p' - \op p = -\frac{b'}{a'}\op P
= -\frac{b}{a}\op P
,
\label{eq330}
\end{align}
each of which corresponds to the {\it input-referred} noise
(Figs. \ref{fig200} and \ref{fig300}).

In the reference \cite{ozawa-pra},
the gain of phase-sensitive amplification in
backaction-evading interaction is properly incorporated
for the definition of error [Eq.~(21a)]
and disturbance [Eq.~(21c)],
but for other cases the gain seems ignored [for example
Eq.~(23c)].

The expectation values of $\op E_*^2$ and $\op D_*^2$ for an
arbitrary state are
\begin{align}
\epsilon_*^2 &= \bracket{\op E_*^2} = \left(\frac{a}{b}\right)^2\sigma^2(\op Q),
\NN
\eta_*^2 &= \bracket{\op D_*^2} = \left(\frac{b}{a}\right)^2\sigma^2(\op P)    
.
\label{eq335}
\end{align}
The Heisenberg uncertainty relation directly follows
\begin{align}
\epsilon_*\eta_* = \sigma(\op Q)\sigma(\op P) \geq \frac{\hbar}{2}
.
\label{eq340}
\end{align}
Now we have found that 
the redefined error $\epsilon_*$ and disturbance $\eta_*$ satisfy
the Heisenberg inequality for $ab\neq0$.

The case $a=0$ can be considered as
the limiting case of $a\rightarrow0$ with $\varDelta$ kept constant.
From Eq.~(\ref{eq335}), we see that for
$\epsilon_*\rightarrow0$, $\eta_*\rightarrow\infty$,
the uncertainty product $\epsilon_*\eta_*$ is conserved and
Eq.~(\ref{eq340}) is always satisfied.
The case of $b=0$ can be treated similarly
as the limit of $b\rightarrow0$.

Thus, revising the definition of error and disturbance
appropriately, we can defend the Heisenberg uncertainty
relation for general class of measurement interaction.
At least, for the linear type interaction (\ref{eq150}) of continuous
variable we can safely use the Heisenberg inequality.

\section{Uncertainty in probability distribution}
One may argue that the new definition of 
$\epsilon_*$ and $\eta_*$ in Eq.~(\ref{eq335}) is questionable
because they diverge 
despite of the finiteness of 
$\sigma(\op Q')$ and $\sigma(\op p')$.
For example, in the limit of $b\rightarrow0$,
the standard deviation of $\op Q$ is finite but $\epsilon_*$ diverges.
Therefore, $\epsilon$, which stays finite as
seen in Eq.~(\ref{eq280}), seems more appropriate.
We will show that this intuitive argument is not correct.

In the uncertainty relations, the second-order moments are used
as the quantitative measures of uncertainties.
In stead here we will use the probability distribution of each variable
in order to accurately examine the effect of interaction.

\subsection{Ideal measurement case}
First we study the case of ideal measurement,
i.e., Type (O).
The initial state for the total system is
\begin{align}
\ket{\varPsi\sub{tot}}=\ket\psi\otimes\ket\varPsi
=\ket\psi\ket\varPsi
,
\label{eq345}
\end{align}
where $\ket\psi$ and $\ket\varPsi$ are the initial states
for the object and probe, respectively.
In the Schr\"odinger picture,
the measurement operator $\op U$ brings the state
into
\begin{align}
\ket{\varPsi\sub{tot}'}=\op U(\ket\psi\ket\varPsi)
.
\label{eq350}
\end{align}

We denote the initial probability distributions for
$\op q$, $\op Q$, $\op p$, and $\op P$ respectively as
\begin{align}
    f(q)&:=|\bracketi{q}{\psi}|^2 = |\psi(q)|^2,
\NN
    F(Q)&:=|\bracketi{Q}{\varPsi}|^2 = |\varPsi(Q)|^2,
\NN
    g(p)&:=|\bracketi{p}{\psi}|^2 = |\phi(p)|^2,
\NN
    G(-P)&:=|\bracketi{P}{\varPsi}|^2 = |\varPhi(P)|^2
.
\label{eq360}
\end{align}

The probability distribution for $\op Q'$ (after the interaction)
is
\begin{align}
F'(Q)=
\infint |\varPsi\sub{tot}'(q, Q)|^2\dd q
,
\label{eq370}
\end{align}
where
\begin{align}
&\varPsi\sub{tot}'(q,Q)
=(\bra q\bra Q)\ket{\varPsi\sub{tot}'}
=\bra q\bra Q\op U\ket{\varPsi\sub{tot}}
\NN
&=\infint\dd q'\infint\dd Q'\bra q\bra Q\op U\ket{q'}\ket{Q'}\bra{q'}\bracketi{Q'}{\varPsi\sub{tot}}
\NN
&=\infint\dd q'\infint\dd Q'
\delta(q-q')\delta(Q-Q'-q')
\psi(q')\varPsi(Q')
\NN
&=\psi(q)\varPsi(Q-q)
\label{eq380}
\end{align}
is the wavefunction of the entire system.
Now we have the change of the distribution as
\begin{align}
F'(Q)
=\infint\dd q f(q)F(Q-q)
= (f*F)(Q)
,
\label{eq390}
\end{align}
where
$
(f*g)(x)=\infint\dd y f(y)g(x-y) %=\infint\dd y f(x-y)g(y)
$    
represents the convolution integral.

The distribution of probe position $\op Q'$ after the
interaction, $F'(Q)$, is a convolution 
of the initial distribution of the object position
$f(q)=|\psi(q)|^2$ with that of the probe position
$F(Q)=|\varPsi(Q)|^2$.

If $F(Q)$ is a sharp function (close to the delta function),
then the initial distribution $f(q)$ can be reproduced.
On the other hand, if $F(Q)$ is a broad function,
the distribution is blurred.
Namely, the width of $F(Q)$ determines the accuracy of
the measurement.

The distribution of object momentum $\op p'$ after the
interaction is
\begin{align}
g'(p)=
\infint|\varPhi\sub{tot}'(p, P)|^2 \dd P
,
\label{eq410}
\end{align}
where
\begin{align}
&\varPhi\sub{tot}'(p,P)
=
(\bra p\bra P)\ket{\varPsi\sub{tot}'}
=
(\bra p\bra P)\op U\ket{\varPsi\sub{tot}}
\NN
&=\infint\dd p'\infint\dd P'\bra p\bra P\op U\ket{p'}\ket{P'}\bra{p'}\bracketi{P'}{\varPsi\sub{tot}}
\NN
&=\infint\dd p'\infint\dd P'
\delta(p-p'+P')\delta(P-P')
\phi(p')\varPhi(P')
\NN
&=\phi(p+P)\varPhi(P)
\label{eq420}
\end{align}
is the wavefunction represented by the momentum basis.
Thus, we have
\begin{align}
g'(p)
=\infint\dd P\, g(p+P)G(-P)=(g*G)(p)
,
\label{eq430}
\end{align}
which is essentially the
convolution of the initial distribution $g(p)$ with
the distribution $G(P)$ of the probe momentum $\op P$.
For a narrow $G(P)$, the distribution $g(p)$ is conserved.
On the other hand, for a wide $G(P)$, $g(p)$ is destructed.
The width of $G(P)$ corresponds to the
strength of disturbance.

Because
$\varPsi(Q)$ and 
$\varPhi(P)$ are the Fourier-transform pair:
\begin{align}
\varPhi(P)&=\bracketi{P}{\varPsi}
=\infint\dd Q \bracketi{P}{Q}\bracketi{Q}{\varPsi}
\NN
&=\frac{1}{\sqrt{2\pi\hbar}}\infint \dd Q \varPsi(Q)\ee^{-\ii QP/\hbar}
,
\label{eq440}
\end{align}
it is impossible to reduce the widths of
$F(Q)=|\varPsi(Q)|^2$ and 
$G(-P)=|\varPhi(P)|^2$, simultaneously.

We represent the first order moment and the variance
of a distribution $f(\cdot)$ as
\begin{align}
m(f)&=\infint q f(q)\dd q,
\NN
\sigma^2(f)&=\infint (q-m(f))^2 f(q)\dd q
,    
\end{align}
respectively.
Using
$\sigma^2(f*g)=\sigma^2(f)+\sigma^2(g)$,
from Eqs.~(\ref{eq390}) and (\ref{eq430}),
we obtain
\begin{align}
    \sigma^2(F')&=\sigma^2(f) + \sigma^2(F),
\NN
    \sigma^2(g')&=\sigma^2(g) + \sigma^2(G)
.
\label{eq450}
\end{align}
These increments in the variances correspond to  $\epsilon^2$ and $\eta^2$
and we have again the uncertainty relation
(\ref{eq40}) 
\begin{align}
\epsilon^2\eta^2&=\left[\sigma^2(F')-\sigma^2(f)\right]
\left[\sigma^2(g')-\sigma^2(g)\right]
\NN
&=\sigma^2(F)\sigma^2(G)\geq\frac{\hbar^2}{4}
.
\label{eq460}
\end{align}
for the ideal measurement.

\subsection{General cases}
For the general transformation (\ref{eq150}),
the probability distributions of $\op Q'$ and $\op p'$ are
\begin{align}
F'(Q)&=\frac{1}{\varDelta}\infint f(-c'Q + a'q)F(d'Q - b'q)\dd q,
\NN
g'(p)&=\varDelta\cdot\infint g(dp + bP)G(-cp - aP)\dd P
.
\label{eq480}
\end{align}
In the case of $ab\neq0$,
these can be written as
\begin{align}
F'(Q)&=\left(f_{1/b'} * F_{1/a'}\right)_{1/\varDelta}(Q),
\NN
g'(p)&=\left(g_{1/a} *  G_{1/b}\right)_\varDelta (p),
\label{eq490}
\end{align}
where we define 
\begin{align}
f_k(x):=kf(kx),
\end{align}
for a function $f(x)$ and a real constant $k>0$.
The graph of the function $f_k(\cdot)$ can be obtained
from that of $f(\cdot)$ by stretching horizontally 
by factor $1/k$ and vertically by factor $k$.
We note that the area is conserved;
\begin{align}
\infint f_k(x)\dd x = \infint f(x)\dd x
,    
\end{align}
and also note
$m(f_k)=(1/k)m(f)$ and
$\sigma^2(f_k)=(1/k^2)\sigma^2(f)$.
The variances of Eq.~(\ref{eq490}) are
\begin{align}
\sigma^2(F')&=b^2\left(\sigma^2(f)+\frac{a^2}{b^2}\sigma^2(F)\right),
\NN
\sigma^2(g')&=a'^2\left(\sigma^2(g)+\frac{b^2}{a^2}\sigma^2(G)\right)
,
\label{eq500}
\end{align}
from which we can regard the error and disturbance as
\begin{align}
\epsilon_*^2=\left(\frac{a}{b}\right)^2\sigma^2(F),
\quad
\eta_*^2=\left(\frac{b}{a}\right)^2\sigma^2(G)
.
\label{eq510}
\end{align}
These are consistent with Eq.~(\ref{eq335}).
Again we have the uncertainty relation for the case of
$ab\neq0$ as
\begin{align}
\epsilon_*\eta_*=\sigma(F)\sigma(G)=\sigma(\op Q)\sigma(\op P)\geq\frac{\hbar}{2}
.
\label{eq520}
\end{align}

For the case of $a=0$ or $b=0$,
Eq.~(\ref{eq480}) simply becomes
\begin{align}
(a=0)\quad&\quad F'(Q)=f(Q), \quad g'(p)=G(p),
\NN
(b=0)\quad&\quad F'(Q)=F(Q), \quad g'(p)=g(p)
.
\label{eq530}
\end{align}
For the case $a=0$, the distribution of $\op q$
is faithfully transferred to that for $\op Q'$ and
therefore no errors creep in; $\epsilon_*=0$.
On the other hand, the distribution of $\op p'$ is 
replaced with that of $\op P$, which contains no information
on $\op p$.
This situation can be considered that the information
is completely destroyed with infinitely large disturbance,
$\eta_*=\infty$.

For the case $b=0$, we have infinitely large error:
$\epsilon_*=\infty$, because no information is
transferred from $\op q$ to $\op Q'$.
On the other hand no disturbances is applied, $\eta_*=0$, because
$\op p$ is conserved.

\subsection{Infinitely large error and disturbance}
In order to clarify further the meaning of infinitely large errors
and disturbances,
let us consider the case $a=0$ as 
a limit of $a\rightarrow0$ for Eq.~(\ref{eq490}).

First we remember that the delta function
can be defined as a limit of parameterized functions
\begin{align}
    f_{1/a}(x)=\frac{1}{a}f(x/a)\rightarrow
        \delta(x) \quad (a\rightarrow0)
\label{eq540}
\end{align}
for an arbitrary function $f(x)$ with unit area.

For 
$b=1$, $a\rightarrow0$,
the first equation of (\ref{eq490}) becomes
\begin{align}
F'(Q)&=\left(f_\varDelta * F_{\varDelta/a}\right)_{1/\varDelta}(Q)
=(f*F_{1/a})(Q)
\NN
&\rightarrow 
(f*\delta)(Q)=f(Q)
\quad(a\rightarrow0)
.
\label{eq550}
\end{align}
We confirm that
$F'(Q)$ is an exact copy of $f(q)$ and
the error-free ($\epsilon_*=0$) measurement is achieved.

The second equation of (\ref{eq490}) with $b=1$ approaches
\begin{align}
g'(p)&=\left(g_{1/a}*G\right)_\varDelta(p)
\NN
&\rightarrow (\delta*G)_\varDelta(p)
=G_\varDelta(p)
\quad (a\rightarrow0),
\label{eq560}
\end{align}
where the original distribution $g(p)$ approaches
the delta function and its shape is lost completely.
This fact justifies the infinitely large disturbance
($\eta_*=\infty$) despite of the finite width of $g'(p)$.

For another way of understanding, Eq.~(\ref{eq560}) can be
rewritten as
\begin{align}
g'(p)=
\left(g * G_a\right)_{1/a'}(p)
\rightarrow G_\varDelta(p)
\quad(a\rightarrow0)
.
\label{eq570}
\end{align}
This equation can be understood as follows:
the original distribution $g(\cdot)$ is convoluted
with $G_{a}(\cdot)$, whose width is infinitely large
or scales as $1/a$.
The resultant distribution $(g*G_{a})(\cdot)$ has
infinitely large width. 
However, the rescaling with $1/a'$ 
results in the finite-width distribution
$G_\varDelta(\cdot)$.
This equation helps to remove the common misunderstanding
that the finite width implies the finite disturbance or
equivalently that the infinite disturbance implies
the infinite width.

Thus, by considering the cases $a=0$ and $b=0$ as
limits of cases $ab\neq0$, we have confirmed that
the Heisenberg uncertainty relation (\ref{eq340}) holds
also in these cases.

\section{Conclusion}

If we borrow the definitions of the error and disturbance
from the ideal measurement case,
the Heisenberg inequality is apparently violated for generalized measurements.
A loosened bound posed by Ozawa is one of the methods for remedy.
Here we have revised the definition of the error and disturbance
considering the gain of generalized measurement interaction.
With these new measures, the validity of the Heisenberg inequality 
is recovered.

Examining the changes of distribution functions caused by the
general measurement interaction,
the physical meanings of infinitely large errors and disturbances are
clarified.

We note that infinitely small, or no errors or disturbances
associated with finite widths, which are just the dual cases, are normally admitted.

In conclusion, with the proper definition for the error and disturbance,
we may not need to discard the Heisenberg inequality for the measurement
at least within the framework of the continuous linear interactions.

\section*{acknowledgment}
The author thanks S. Tamate and T. Nakanishi for carefully reading
the manuscript.
This work is supported through
the global COE program
``Photonics and Electronics Science and Engineering,''
at Kyoto University
by the Ministry of Education, Culture, Sports, Science, and Technology
of Japan.

\appendix*
\section{Interaction for measurement}
We consider the interaction $\op U$ which provides
the linear relation between the object 
and probe positions.
It transforms the basis ket $\ket q\ket Q$ as
\begin{align}
\op U:\,
\ket q\ket Q \mapsto \ket{q'}\ket{Q'}=\omega\ket{dq+cQ}\ket{aQ+bq}
,
\label{eq580}
\end{align}
where $a, b, c, d\in\mathbb R$ are constant and
$\omega$ will be determined from the unitary condition.

In the position basis,
$\op U$ is represented as 
\begin{align}
\op U
&=\infint\!\infint\!\infint\!\infint\dd q'\dd q\dd Q'\dd Q
\NN
&\hspace*{3em}
U(q', Q', q, Q)\ket{q'}\ket{Q'}\bra{q}\bra{Q}    
.
\label{eq590}
\end{align}
In order to satisfy Eq.~(\ref{eq580}),
the matrix element should be of the following form:
\begin{align}
&U(q', Q', q, Q) = 
\bra{q'}\bra{Q'}\op U\ket q\ket Q
\NN
&=\omega\delta(q' - dq - cQ)\delta(Q' - aQ - bq)
,
\label{eq600}
\end{align}
where $\delta(\cdot)$ is Dirac's delta function.
Substitution of (\ref{eq600}) into the unitary condition:
\begin{align}
&\infint\infint
\dd q'\dd Q'
\,U^*(q', Q', q'', Q'')U(q', Q', q, Q)
\NN
&=\delta(q''-q)\delta(Q''-Q)    
,
\label{eq610}
\end{align}
yields
\begin{align}
&\infint\!\infint\dd q'\dd Q'
|\omega|^2\delta(q'\!-dq''\!-cQ'')\delta(Q'\!-aQ''\!-bq'')
\NN
&\hspace*{5em}
\times\delta(q'\!-dq-cQ)\delta(Q'\!-aQ-bq)
\NN
&=
|\omega|^2\delta(d(q''\!-q)+c(Q''\!-Q))\delta(a(Q''\!-Q)+b(q''\!-q))
\NN
&=
\frac{|\omega|^2}{|ad-bc|}\delta(q''\!-q)\delta(Q''\!-Q)
,
\label{eq620}
\end{align}
where we have used the formula:
\begin{align}
\delta(ax+by)\delta(cx+dy)=\delta(x)\delta(y)/|ad-bc|
.
\label{eq622}
\end{align}
Thus the unitary condition is given as
\begin{align}
\omega=\sqrt{|\varDelta|}, \quad
\varDelta=ad - bc\,(\neq0)
.
\end{align}

For the unitary operator which is generated continuously
with a Hamiltonian, $\varDelta>0$ can be assumed
because the identity operator, $a=d=1$, $b=c=0$,
satisfies $\varDelta=1>0$.
The sign of $a$ can be inverted by changing the sign of $\op Q$ and $\op q'$.
The sign of $b$ can also be inverted with $\op q$ and $\op q'$.
Therefore, without loss of generality we can assume $a\geq0$ and $b\geq0$.

The momentum presentation of $\op U$ is
\begin{align}
&
V(p', P', p, P)=
\bra{p'}\bra{P'}\op U\ket p\ket P
\NN
&=\infint\!\infint\!\infint\!\infint\dd q'\dd q\dd Q'\dd Q
\NN
&\hspace*{5em}
U(q',Q',q,Q)
\bracketi{p'}{q'}\bracketi{P'}{Q'}\bracketi{q}{p}\bracketi{Q}{P}
\NN
&=\infint\!\infint\!\infint\!\infint\dd q'\dd q\dd Q'\dd Q
\NN
&\hspace*{5em}
U(q',Q',q,Q)
\frac{\ee^{\ii(pq-p'q')/\hbar}}{2\pi\hbar}
\frac{\ee^{\ii(PQ-P'Q')/\hbar}}{2\pi\hbar}
\NN
&=\frac{\omega}{(2\pi\hbar)^2}
\infint\!\infint\dd q\dd Q
%\,\ee^{\ii(p-dp'-bP')q/\hbar} \ee^{\ii(P-aP'-cp')Q/\hbar}
\,\ee^{\ii[(p-dp'-bP')q+(P-aP'-cp')Q]/\hbar}
\NN
&=
\omega\delta(p-dp'-bP')\delta(P-aP'-cp')
\NN
&=
\omega^{-1}\delta(p'-a'p+b'P)\delta(P'-d'P+c'p)
,
\label{eq630}
\end{align}
where
$a'=a/\varDelta$, $b'=b/\varDelta$, $c'=c/\varDelta$ and $d'=d/\varDelta$.
We have utilized Eq.~(\ref{eq622}).
The eigenket of momenta, $\ket p\ket P$, is transformed as
\begin{align}
\op U:\,
\ket p\ket P \mapsto \frac{1}{\sqrt\varDelta}\ket{a'p-b'P}\ket{{-c'}p+d'P}
.
\label{eq640}
\end{align}

From Eqs.~(\ref{eq580}) and (\ref{eq640}), we can confirm the 
commutation relation:
\begin{align}
[\op Q', \op P']
&=[a\op Q + b\op q,\, -c'\op p + d'\op P]
\NN
&=ad'[\op Q, \op P] - bc'[\op q, \op p]
=\ii\hbar\op 1
.
\label{eq650}
\end{align}

\end{document}